\documentstyle[preprint,revtex]{aps} 
\begin{document}
\draft
\preprint{August 1996}
\begin{title}
$SU(m|n)$ supersymmetric Calogero-Sutherland model \\
confined in harmonic potential 
\end{title}
\author{C.-A. Piguet, D. F. Wang and C. Gruber}
\begin{instit} 
Institut de Physique Th\'eorique\\
Ecole Polytechnique F\'ed\'erale de Lausanne\\
PHB-Ecublens, CH-1015 Lausanne, Switzerland
\end{instit}
\begin{abstract}
In this work, we study a continuous quantum system of a mixture of
bosons and fermions with the supersymmetry $SU(m|n)$. The particles 
are confined in a harmonic well and interact with each other
through the $1/r^2$ interaction. The ground state wavefunction 
is constructed explicitly for the most general $SU(m|n)$ case, 
with the ground state energy given explicitly. The full energy spectrum
of excitations in the $SU(m|n)$ model is also equal spaced. 
In the limiting case where there are no bosons in the system,
our results reduce to those obtained previously. 
\end{abstract}
\narrowtext

There have been considerable recent interests in study of the Calogero-Sutherland
models\cite{calogero,sutherland}, as well as their various generalizations. 
One of the examples is the multi-component generalization of the CS
model. For the $SU(n)$ fermionic particles confined in a harmonic oscillator
potential and interacting with each other through the $1/r^2$ interaction,
the ground state wavefunction was constructed explicitly, and the ground 
state energy was found in previous work\cite{vacek}. The excitation of the fermionic system was 
found to be equal spaced\cite{ka1,gruber,poly,vacek,ya}.  
On the other hand, the system 
consisting of bosons with $SU(n)$ spins has been studied for the 
pairwise interaction $\lambda (\lambda+P_{ij}^\sigma)/x_{ij}^2$ 
in presence of harmonic potential
confinement, where $P_{ij}^\sigma$ permutes the spins of the bosons.  

In the following, we will investigate the ground state wavefunction of the 
the most general supersymmetric $SU(m|n)$
multi-component CS model confined in a harmonic potential. In this case,
the system consists of a mixture of bosons and fermions. The ground state
wavefunction for this supersymmetric case has been unknown by far. 
In this work, we provide the ground state wavefunction, and compute 
its eigenenergy explicitly in the subspace where the number of particles 
of each flavor is fixed. In the limiting case where there are only fermions
in the system, our results reduce to those obtained by previous authors.  

We consider a one-dimensional quantum system of $Q$ interacting particles 
confined in a harmonic potential of frequency $\omega$. 
The hamiltonian under consideration reads
\begin{equation}
H=-\frac{1}{2}\sum_{i=1}^{Q}\frac{\partial^2}{\partial q_{i}^{2}}+
\frac{\omega^2}{2}\sum_{i=1}^{Q}q_{i}^2
+\sum_{j>i}\frac{\lambda(\lambda-Y_{ij})}{(q_{i}-q_{j})^2}
\label{ham}
\end{equation}
where $q_{1},\ldots,q_{Q}$ are the coordinates of the particles and $Y_{ij}$ is
the operator that exchanges the positions of particles $i$ and $j$. The real
constant $\lambda$ is supposed to be greater than $1/2$. This hamiltonian is
exactly the one considered in \cite{vacek} for a system of only fermions. 

We now specify to the case where we have $M$ fermions and $N$ bosons $(M+N=Q)$ with 
$SU(m|n)$ spin degrees of freedom.
We denote $x_1,\ldots,x_M$ the positions of the fermions, $y_1,\ldots,y_N$
the positions of the bosons, $\sigma_1,\ldots,\sigma_M$ the spins of the fermions and
$\tau_1,\ldots,\tau_N$ the spins of the bosons. We 
also use the notation that $(q_1,q_2,\cdots,q_Q)=(x_1,\cdots,x_M|y_1,y_2,\cdots,y_N)$.
With these notations, the
hamiltonian Eq.~(\ref{ham}) is made of three parts
\begin{equation}
H=H_F+H_B+H_M
\end{equation}
where 
\begin{eqnarray}
H_F&=&-\frac{1}{2}\sum_{i=1}^{M}\frac{\partial^2}{\partial x_{i}^2}+
\frac{\omega^2}{2}\sum_{i=1}^{M}x_{i}^2+
\sum_{j>i}\frac{\lambda(\lambda-Y_{ij})}{(x_i-x_j)^2}\nonumber\\
H_B&=&-\frac{1}{2}\sum_{\alpha=1}^{N}\frac{\partial^2}{\partial y_{\alpha}^2}+
\frac{\omega^2}{2}\sum_{\alpha=1}^{N}y_{\alpha}^2+
\sum_{\beta>\alpha}\frac{\lambda(\lambda-Y_{\alpha\beta})}{(y_{\alpha}-y_{\beta})^2}\nonumber\\
H_M&=&\sum_{i,\alpha}\frac{\lambda(\lambda-Y_{i\alpha})}{(x_i-y_{\alpha})^2}.
\end{eqnarray}
$H_F$ ($H_B$) is the hamiltonian for a system of $M$ ($N$) interacting fermions
(bosons). $H_M$ represents the interaction between the fermions and the bosons.
 
We now wish to find the ground state wavefunction of the hamiltonian.
For this, we propose a wavefunction which is a product of three terms
\begin{eqnarray}
\Psi(x_1,\sigma_1;\ldots;x_M,\sigma_M|y_1,\tau_1;\ldots;y_N,\tau_N)
&=&\Psi_F(x_1,\sigma_1,\ldots,x_M,\sigma_M)\cdot\Psi_B(y_1,\ldots,y_N)\nonumber\\
\cdot\Psi_M(x_1,\ldots,x_M,y_1,\ldots,y_N)
\label{wf}
\end{eqnarray}
where
\begin{eqnarray}
&&\Psi_F(x_1,\sigma_1;\ldots;x_M,\sigma_M)
=\prod_{j>i}|x_j-x_i|^{\lambda}(x_j-x_i)^{\delta_{\sigma_{i}\sigma_{j}}}
\exp\left(i\frac{\pi}{2}{\rm sgn} (\sigma_{i}-\sigma_{j})\right)\prod_{i}
\exp\left(-\frac{\omega}{2}x_{i}^2\right)\nonumber\\
&&\Psi_B(y_1,\ldots,y_N)=\prod_{\beta>\alpha}|y_\beta-y_\alpha|^{\lambda}
\prod_{\alpha}\exp\left(-\frac{\omega}{2}y_{\alpha}^2\right)\nonumber\\
&&\Psi_M(x_1,\ldots,x_M,y_1,\ldots,y_N)=\prod_{i,\alpha}|x_i-y_{\alpha}|^{\lambda}.
\end{eqnarray}
Let us first remark that this wavefunction has the correct symmetry:
$M_{ij}\Psi=-\Psi$ for the fermions and $M_{\alpha\beta}\Psi=\Psi$ for the
bosons, where $M$ is the operator that exchanges the particles.
We can also remark that the spin of the bosons does not enter this
wavefunction.

Let us now prove that the proposed wavefunction Eq.~(\ref{wf}) is an eigenstate of the hamiltonian.
The wavefunction $\Psi_F$ is the ground state of the system with only fermions
found in \cite{vacek}. We thus have 
\begin{equation}
H_F\Psi_F=\frac{\omega}{2}\left[\lambda
M(M-1)+\sum_{k=1}^{m}M_{k}^2\right]\Psi_F
\end{equation}
where $M_k$ is the number of fermions with spin $k$.
The wavefunction $\Psi_B$ is the ground state of the system with only bosons. It
is easy to check that
\begin{equation}
H_B\Psi_B=\frac{\omega}{2}\left[\lambda
N(N-1)+N\right]\Psi_B.
\end{equation}
With this, one can show that
\begin{eqnarray}
\frac{1}{\Psi}(H_F+H_B)\Psi&=&\frac{\omega}{2}\left[\lambda
\{ M(M-1)+N(N-1)+2MN \} +\sum_{k=1}^{m} M_k^2+ N \right]\nonumber\\
& &-\sum_{i,\alpha}\frac{\lambda(\lambda-1)}
{(x_i-y_{\alpha})^2}-\sum_{\alpha,i\neq
j}\frac{\lambda\delta_{\sigma_{i}\sigma_{j}}}
{(x_i-x_j)(x_i-y_{\alpha})}.
\label{unwant}
\end{eqnarray}
To apply the interaction part on the wavefunction, we first compute the effect
of $Y_{i\alpha}$
\begin{equation}
\frac{Y_{i\alpha}\Psi}{\Psi}=\prod_{j\neq i}
\left(\frac{x_j-y_{\alpha}}{x_j-x_i}\right)^{\delta_{\sigma_{i}\sigma_{j}}}=
\prod_{j\neq i}\left(1+\delta_{\sigma_{i}\sigma_{j}}\frac{x_i-y_{\alpha}}
{x_j-x_i}\right).
\end{equation}
Thus
\begin{equation}
\frac{1}{\Psi}\left[\sum_{i,\alpha}\frac{\lambda(\lambda-K_{i\alpha})}{(x_i-y_{\alpha})^2}
\Psi\right]=\sum_{i,\alpha}\frac{\lambda^2}{(x_i-y_{\alpha})^2}-
\sum_{i,\alpha}\frac{\lambda}{(x_i-y_{\alpha})^2}\prod_{j\neq i}
\left(1+\delta_{\sigma_{i}\sigma_{j}}\frac{x_i-y_{\alpha}}{x_j-x_i}\right)
\label{exp}
\end{equation}
We can expand the second term in power of $x_i-y_{\alpha}$. The terms of order
greater than 1 are zero as will be proved in the following. Using this result, we have
\begin{equation}
\frac{1}{\Psi}\left[\sum_{i,\alpha}\frac{\lambda(\lambda-K_{i\alpha})}{(x_i-y_{\alpha})^2}
\Psi\right]=\sum_{i,\alpha}\frac{\lambda(\lambda-1)}{(x_i-y_{\alpha})^2}-
\sum_{\alpha,i\neq
j}\frac{\lambda\delta_{\sigma_{i}\sigma_{j}}}{(x_i-y_{\alpha})
(x_j-x_i)}.
\end{equation}
These terms cancel with the fermionic and bosonic inconstant terms in
Eq.~(\ref{unwant}) to give the following
ground state energy:
\begin{equation}
E_{G}=\frac{1}{2}\omega\left[\lambda \{ M(M-1)+N(N-1)+2MN \} +\sum_{k=1}^{m}M_{k}^2+N\right].
\label{energy}
\end{equation}

We have now to prove that the terms of order greater than 1 of the second term
of Eq.~(\ref{exp}) are zero.
For this, let us consider a term of order $s$. We have $s+1$ particles with the
same spin with coordinates $i,k_1,\ldots,k_s$. Let us now show that if we sum
over all the permutations $i\leftrightarrow k_{j}$ with $j=1,\ldots,s$ while
keeping $\alpha$ fixed we get a zero contribution:
\begin{equation}
-\lambda\sum_{\{i,k_1,\ldots,k_s\}}(x_i-y_{\alpha})^{s-2}\prod_{j=i}^{s}
\frac{1}{x_{k_j}-x_i}=0.
\end{equation}
For this, following \cite{vacek}, we expand the term $(x_i-y_{\alpha})^{s-2}$ 
and express the product in terms of Vandermonde determinant
\begin{equation}
\prod_{j=i}^{s}\frac{1}{x_{k_j}-x_i}=(-1)^{s}\frac{V^{(s)}(x_{k_1},\ldots,x_{k_s})}
{V^{(s+1)}(x_{k_1},\ldots,x_{k_s},x_i)}
\end{equation}
to get
\begin{eqnarray}
& &-\lambda\sum_{\{i,k_1,\ldots,k_s\}}(x_i-y_{\alpha})^{s-2}\prod_{j=i}^{s}
\frac{1}{x_{k_j}-x_i}\nonumber\\
&=&-\lambda\sum_{\{i,k_1,\ldots,k_s\}}\sum_{t=0}^{s-2}\left(
\begin{array}{c}
s-2\\
t\end{array}\right)
x_{i}^{t}(-1)^{s-2-t}y_{\alpha}^{s-2-t}(-1)^{s}\frac{V^{(s)}(x_{k_1},\ldots,x_{k_s})}
{V^{(s+1)}(x_{k_1},\ldots,x_{k_s},x_i)}\nonumber\\
&=&-\lambda\sum_{t=0}^{s-2}\left(
\begin{array}{c}
s-2\\
t\end{array}\right)
(-1)^{s-2-t}y_{\alpha}^{s-2-t}(-1)^{s}\frac{W^{(s,t)}(x_{j_1},\ldots,x_{j_s},x_i)}
{V^{(s+1)}(x_{k_1},\ldots,x_{k_s},x_i)}
\end{eqnarray}
where
\begin{equation}
W^{(s,t)}(x_{j_1},\ldots,x_{j_s},x_i)={\rm det}\left(\begin{array}{ccccc}
1 & 1 & \ldots & 1 & 1 \\
x_{j_1} & x_{j_2} & \ldots & x_{j_s} & x_i \\
\vdots & \vdots & & \vdots & \vdots \\
x_{j_1}^{s-1} & x_{j_2}^{s-1} & \ldots & x_{j_s}^{s-1} & x_{i}^{s-1}\\ 
x_{j_1}^{t} & x_{j_2}^{t} & \ldots & x_{j_s}^{t} & x_{i}^{t}\\
\end{array}\right). 
\end{equation}
We can now conclude, remarking that since $0\leq t\leq s-2$ two rows of the det
$W$ are the same and thus the whole expression is zero. We have thus proved that
the wavefunction Eq.~(\ref{wf}) is an eigenstate of the hamiltonian Eq.~(\ref{ham}) with the
energy Eq.~(\ref{energy}) in the case of a mixture of fermions and bosons with
$SU(m|n)$ spin symmetry. 

We anticipate this wavefunction to be the ground state of the system in 
the supersymmetric case. First of all, without bosons, our wavefunction
will reduce to the ground state wavefunction for the multi-component fermionic
system studied before\cite{vacek}. Second, when both $M$ and $N$ are nonzero, it is impossible
to construct lower energy eigenstates by using the lowering operators, in the same
way as for the non-uniform supersymmetric t-J model\cite{gruber,liu}.
There are $M+N$ lowering operators, $a_i, i=1, 2\cdots, Q$\cite{poly}. In fact, 
we have checked explicitly the following relations: 
\begin{eqnarray}
&&(\sum_{i=1}^M a_i)|\Psi>=0,
~~(\sum_{i=1}^M a_i \sigma_i^z) |\Psi>=0,\nonumber\\ 
&&(\sum_{i=M+1}^Q a_i) |\Psi>=0,
~~(\sum_{i=M+1}^Q a_i \tau_{i-M}^z)|\Psi>=0,
\end{eqnarray}
where $\Psi$ is the wavefunction that we constructed here for the 
supersymmetric system.  
These two evidences support our conclusion that the wavefunction
we have constructed for the supersymmetric $SU(m|n)$ case is 
indeed the ground state in the subspace where the number of particles
of each flavor is fixed.

Having studied the ground state wavefunction, we can now in fact
construct a set of non-trivial excited states.
These states are obtained by  multiplying the ground state wavefunction by a
symmetrical product $F$ of Hermite polynomials. Without bosons, the following
excited states will reduce to those obtained before\cite{vacek}. Our
wavefunctions take the form of
\begin{equation}
\Phi=\Psi\cdot F=\Psi\cdot\sum_{m_1,\ldots,m_Q\atop{m_1+\ldots+m_Q=I}}\prod_{i=1}^{Q}
\frac{1}{m_{i}!}H_{m_i}(\sqrt{\omega}q_i)
\end{equation}
where $\Psi$ is the ground state given by Eq.(\ref{wf}) and the $m_i$ are
positive integers whose sum is the quantum number $I$.
To prove that this wave functon is an eigenfunction of our hamiltonian, we
simply use the fact that $\Psi$ is the ground state to get the following
eigenequation
\begin{eqnarray}
\frac{H\Phi}{\Phi}&=&E_{G}+\omega
I-{1\over F} \sum_{i,j=1\atop{i>j}}^{M}\frac{\lambda+\delta_{\sigma_{i}
\sigma_{j}}}{q_i-q_j}(\frac{\partial F}{\partial q_i}-\frac{\partial F}{\partial q_j})
-{1\over F} \sum_{i,j=M+1\atop{i>j}}^{Q}\frac{\lambda}
{q_i-q_j}(\frac{\partial F}{\partial q_i}-\frac{\partial F}{\partial q_j})\nonumber\\
& &-{1\over F} \sum_{i=1}^{M}\sum_{j=M+1}^{Q}\frac{\lambda}
{q_i-q_j}(\frac{\partial F}{\partial q_i}-\frac{\partial F}{\partial q_j}).
\end{eqnarray}
The three last terms are zero since we have 
\begin{equation}
\frac{\partial F}{\partial q_i}=\frac{\partial F}{\partial q_j}
\end{equation}
for all $i$ and $j$. The corresponding energy spectrum is thus equal spaced and given by
\begin{equation}
E=E_{G}+\omega I.
\end{equation}
These wavefunctions of excited states are not the full set of excited states
of the system. However, these states have already spanned the full energy levels
for the system, with $I=0,1,2,\cdots,+\infty$. 

In summary, we have considered the supersymmetric long range model 
confined in a harmonic potential. The ground state wavefunction and 
the ground state energy have been provided by us. We have also 
constructed a set of non-trivial excited states of Jastrow form.
In the limit case of no bosons, our results reduce to those
obtained previously for the $SU(n)$ fermionic gas confined in 
harmonic potential.  
For the supersymmetric model studied here, the system still has lowering
and raising operators, and the full energy spectrum of this system is also equal spaced.
In fact, assuming that the degeneracy of each energy level is the
same as if there were no interaction between the particles, one can further
construct the thermodynamics of the system very easily.  

This work was supported in part by the Swiss National Science Foundation. 



\begin{references} 
\bibitem{calogero}
F. Calogero, J. Math. Phys. {\bf 10}, 2191, 2197 (1969); {\it ibid.}, {\bf 12},
419 (1971).
\bibitem{sutherland}
B. Sutherland, J. Math. Phys. {\bf 12}, 246,251 (1971); Phys. Rev. {\bf A 4},
2019 (1971); {\it ibid.} {\bf A 5}, 1372 (1971).
\bibitem{ka1} N. Kawakami and N. Kuramoto, Phys. Rev. {\bf B 50}, 4664 (1994). 
\bibitem{gruber} C. Gruber and D. F. Wang, Phys. Rev. {\bf B 50}, 3103 (1994). 
\bibitem{poly}
A. P. Polychronakos, Phys. Rev. Lett. {\bf 69}, 703 (1992);
L. Brink, T. H. Hansson and M. A. Vasiliev,
Phys. Lett. {\bf B 286}, 109 (1992).
\bibitem{vacek}
K. Vacek, A. Okiji and N. Kawakami, Phys.Rev. {\bf B 49}, 4637 (1992);
J. Phys. {\bf A 7}, L201 (1994).
\bibitem{liu} J. T. Liu and D. F. Wang, 
{\tt cond-mat/9602093}, appearing in Int. J. Mod. Phys. {\bf B}, (1996).
\bibitem{ya} T. Yamamoto and O. Tsuchiya, 
J. Phys. {\bf A 29}, 3977 (1996).
\end{references}
\end{document}